\documentclass[a4paper,12pt]{article}
\usepackage{amsmath}
\usepackage{epsf}
\usepackage{comment}

\usepackage{natbib}
\bibpunct[, ]{(}{)}{;}{u}{}{,}

\usepackage{asp2004_DoG}

\newcommand{\be}{\begin{equation}}
\newcommand{\ee}{\end{equation}}

\newcommand{\ba}{\begin{array}}
\newcommand{\ea}{\end{array}}

\newcommand{\bea}{\begin{eqnarray}}
\newcommand{\eea}{\end{eqnarray}}

\newcommand{\bear}{\begin{eqnarray*}}
\newcommand{\eear}{\end{eqnarray*}}

\newcommand{\bn}{\begin{enumerate}}
\newcommand{\en}{\end{enumerate}}

\newcommand{\bi}{\begin{itemize}}
\newcommand{\ei}{\end{itemize}}

\newcommand{\bl}{\begin{list}}
\newcommand{\el}{\end{list}}

\newcommand{\bvb}{\begin{verbatim}}	
\newcommand{\evb}{\end{verbatim}}	

\newcommand{\bt}{\begin{tabular}}
\newcommand{\et}{\end{tabular}}

\newcommand{\bc}{\begin{center}}
\newcommand{\ec}{\end{center}}

\newcommand{\ed}{\end{document}}

\newcommand{\bfr}{\begin{flushright}}
\newcommand{\efr}{\end{flushright}}

\newcommand{\bfl}{\begin{flushleft}}
\newcommand{\efl}{\end{flushleft}}

\newcommand{\PBS}[1]{\let\temp=\\#1\let\\=\temp}

\newcounter{RimNumber}

\newcommand{\rim}[1]{{\sl\rm\uppercase\expandafter{\romannumeral#1}}}

\DeclareMathOperator{\msin}{\smash[t]{\mathrm{sin}}}

\begin{document}
\sloppy 

\begin{center}
\LARGE\bf
On a Source of Systematic Error\\ in Absolute Measurement of Galactocentric
Distance from Solving for the Stellar Orbit Around Sgr A* 
\end{center}
\thispagestyle{empty}
\sc

\vskip 12pt
\begin{center}
\large
Igor' I. Nikiforov%
\end{center}
\it

\vskip 1pt
\begin{center}
Sobolev Astronomical Institute, St.~Petersburg State
University, Universitetskij pr.~28,  Staryj Peterhof, St.~Petersburg 198504,
Russia, nii@astro.spbu.ru
\end{center}
\rm

\begin{abstract}
\citet{Eea03,Eea05} derived absolute (geometrical) estimates of the distance
to the center of the Galaxy, $R_0$, from the star S2 orbit around Sgr~A* on
the assumption that the intrinsic velocity of Sgr~A* is negligible. This
assumption produces the source of systematic error in $R_0$ value owing to a
probable motion of Sgr~A* relative to the accepted velocity reference system
which is arbitrary to some extent. \citeauthor{Eea05}\  justify neglecting all
three spatial velocity components of Sgr~A*  mainly by low limits of Sgr~A*'s
proper motion of 20--60 km/s. In this brief paper, a simple analysis in the
context of the Keplerian dynamics was used to demonstrate that neglect of even
low (perhaps, formal) radial velocity of Sgr~A* leads to a substantial
systematic error in $R_0$: the same limits of 20--60~km/s result in $R_0$
errors of 1.3--5.6\%, i.e., (0.1--0.45)$\times (R_0/8)$~kpc, for current S2
velocities. Similar values for Sgr~A*'s tangential motion can multiply this
systematic error in the case of S2 orbit by factor ${\approx}1.5$--$1.9$ in the limiting
cases.  
\end{abstract}

\section{Introduction}

The distance from the Sun to the center of the Milky Way, $R_0$, is a
fundamental Galactic constant for solving many astronomical and
astrophysical problems \citep[see, e.g.,][]{R93}. That is why, in its
turn, the problem of determination of $R_0$ remains topical over many
years. Absolute (i.e., not using luminosity calibrations) estimates of
$R_0$ with a current 3\% formal uncertainty from modelling the star S2 orbit
around the compact concentration of dark mass, the so-called
``supermassive black hole'', associated with the radio source Sgr~A*
\citep{Eea03,Eea05,Tea06} present a major breakthrough in measuring
$R_0$! (For brevity, from here on the object in focus of S2 orbit will be
referred to as ``Sgr~A*''.)

However, even though to take no notice the issue on coincidence of Sgr~A*
with the dynamical and/or luminous center(s) of our Galaxy \citep[see
discussion in][]{Nishiyama_ea06}, taken alone the modelling the orbital motion
of a star near Sgr~A* can be plagued with various {\em systematic\/} sources
of error. Since \citeauthor{Eea05}\  solved for the {\em Keplerian\/} orbit of
the star S2, in the literature {\em relativistic\/} effects and {\em non-Keplerian\/}
orbit modelling are primarily explored for this problem
\citep[e.g.,][]{Eea05,Mouawad_ea05,Weinberg_ea05}. 

Meanwhile, \citeauthor{Eea05}\ also used another {\em assumption that the
intrinsic velocity of Sgr~A* is negligible\/}. This assumption can produce
the source of systematic error in $R_0$ value owing to a probable motion of
Sgr~A* relative to the accepted velocity reference system which is
arbitrary to some extent.  Thus far, no consideration has been given to the
role of this factor in measuring $R_0$.

In this study, a simple analysis  is used to evaluate the {\em impact of an
unaccounted motion of Sgr~A*\/} (i.e., the focus of S2 orbit) {\em on an $R_0$
value\/} found from the formal solution of orbit. The Keplerian dynamics only is
taken into consideration because relativistic and non-Keplerian effects seem
to be insignificant for measuring $R_0$
\citep{Eea05,Mouawad_ea05,Weinberg_ea05}. Particular attention has been given
to the impact of a nonzero radial velocity of Sgr~A* relative to the
Local Standard of Rest.

\section{Structure of the Problem on Determination of Orbital Parameters,
Distance to and Mass at Orbital Focus (Sgr~A*)}

The completeness of solution of the problem in question is determined by the
type of available data on motion of an individual star (S2).

\subsection{Star's Proper Motions Alone are Available}%

In this case, {\em all six orbital parameters are solved, except that only the
absolute value of the inclination angle, $i$, is determined\/}, leaving the
questions of the direction of revolution (prograde, $i>0$, or retrograde,
$i<0$) and where along the line of sight the star is located behind the
central object unresolved
\citep[e.g.,][]{Ghez_ea03}. Besides, {\em the semimajor axis is derived in
angular units\/} (in arcsec), hereafter $a''$. The distance to the focus,
i.e., $R_0$, and the central mass, $M$, can not be solved.

With accepted $R_0$, however,  the value of semimajor axis, $a$, is calculated
in linear units (in kpc) and the central mass is found from Kepler's third law
\be\label{M}
M=n^2a^3/G, \qquad n=2\pi/P,
\ee
where $G$ is the gravitational constant, $n$ is the mean motion, and $P$ is
the orbital period, as it has been done in \citet{Schoedel_ea02}.

\subsection{Proper Motions and at Least a Single Measurement of Radial
Velocity of Star are Available}%

In this case, {\em the problem is completely solved\/} if the value of star's
radial (line-of-sight) velocity, $V_r$, is significantly different from zero (more exactly,
from the radial velocity of the focus).

A. {\em The sign of\/} $V_r$ {\em determines the sign of $i$.\/} Consequently,
this also breaks the ambiguity in the direction of rotation and in star's
location along  the line of sight relative to the focus
\citep[e.g.,][]{Ghez_ea03}. 

B. {\em The absolute value of\/} $V_r$ {\em determines values\/} $R_0$ {\em
and\/} $M$. To gain greater insight into the fact of the matter, the problem
can be symbolically divided into two subproblems: (1) the determination of
orbital parameters from the proper motions alone and (2) the determination,
knowing the orbit, of the distance to focus ($R_0$) and of the central mass from
the measurement(s) of $V_r$.  These subproblems are almost independent in the
case of modelling the motion of stars around Sgr~A*, since up to now proper
motion measurements are numerous, but $V_r$ ones are few or at all $V_r$
actually is single, for any S star with solved orbit. So, $V_r$ measurement(s)
contribute(s) almost nothing to the knowledge of orbit, and vice versa proper
motion measurements do not directly determine neither $R_0$ nor $M$. Thus,
such breaking the problem down seems to be quite realistic.

If so, the value of $|V_r|$ may be considered as determining $R_0$ and
$M$ from known orbital parameters as follows.

({\bf i}) The orbit elements enable to find the ratio between $|V_r|$ and the total
space velocity, $V$, for the moment $t$: 
\be\label{Vr/V}
V_r^2/V^2=\frac{[e\sin v\sin u +(1+e\cos v)\cos u]^2\msin^2i}{1+2e\cos v+e^2},
\ee
where $e$ is the eccentricity, $v$ is the true anomaly, $u=v+\omega$\/ is
the argument of latitude, $\omega$ is the argument of pericenter. A
value of $v$ can be calculated from classical formalism:  $$
\tan(v/2)=\sqrt{(1+e)/(1-e)}\tan(E/2),
$$
$$
E-e\sin E={\cal M},\qquad {\cal M}=n(t-t_0)+{\cal M}_0,
$$
where $E$ and $\cal M$ are the eccentric and mean anomalies, correspondingly 
\citep[e.g.,][]{Subbotin68}. Consequently, the knowledge of $|V_r|$ determines $V$.

({\bf ii}) The value of total velocity $V$ can be expressed as
\be\label{V}
V=na\left(\frac{1+2e\cos v+e^2}{1-e^2}\right)^{1/2}.
\ee
From this equation, the value of $a$ {\em in linear units\/} can be
calculated. Then the ratio between $a$ values in linear and angular units
gives $R_0$:
\be\label{R0}
R_0=\frac{a\text{ [kpc]}}{a''}.
\ee

({\bf iii}) Using Eq.~(\ref{M}) with $a$ in linear units determines the
central mass $M$.

\section{Systematic Error in $\mathbf{R_0}$ Owing to a Nonzero Motion of Orbital Focus
(Sgr~A*)}

\subsection{Nonzero Radial Velocity of Sgr~A*}\label{Vr_ne_0}

\citet{Eea03,Eea05} assume that the radial velocity of Sgr~A*,
$V_r^*\equiv V_r(\text{Sgr A*})$, relative to the Local Standard of Rest (LSR) is
zero. Neglect of a possible radial motion of Sgr~A* is equivalent to the
introducing a corresponding systematic error in all $V_r$ values. This
error is equal to a value of $V_r^*$ and is the same in
all measurements of $V_r$.  From Eqs.~(\ref{Vr/V})--(\ref{R0}) follows
that the relative systematic error in $V_r$ velocity fully converts to
the relative systematic error in $R_0$, i.e., 
\be\label{delta}
\delta_{\text{sys}}\equiv\frac{\sigma_{\text{sys}}(V_r)}{|V_r|}=
\frac{\sigma_{\text{sys}}(R_0)}{R_0}.
\ee

These simple considerations make it possible readily to evaluate the
systematic error in $R_0$ knowing typical values of $V_r$ used for the
determination of  distance to S2/Sgr~A*. The first S2 radial velocity
measurement of $V_r=-510\pm 40$~km/s by \cite{Ghez_ea03} was obtained just 30
days after the star's passage through the pericenter point when $V_r$  was
changing very rapidly. Therefore, this measurement contributes to the solution
for $R_0$ much less then subsequent ones, hence the evaluation of
$\sigma_{\text{sys}}(R_0)$ must lean upon these latter. Besides, the
subsequent radial velocities, having substantially higher absolute values, 
give a {\em lower\/} limit for $\sigma_{\text{sys}}(R_0)$. 

\citeauthor{Eea05}\ justify neglecting all  three spatial velocity components
of Sgr~A*  mainly by low limits of Sgr~A*'s proper motion of 20--60 km/s
\citep{Eea05}. Such values of radial velocities seem to be quite plausible for
massive objects in the Galactic center \cite[see][]{Blitz94}.
Table~\ref{tab_r0sys} presents values of systematical errors in $R_0$
calculated for possible Sgr~A*'s radial velocities of $V_r^*=20$ and 60~km/s
with $R_0=7.5$ and 8.0~kpc \citep{R93,N04,Tea06}. In Table~1, $\langle
V_r\rangle$ is the average of  velocities $V_r$, used for estimation of $R_0$
in \cite{Eea05}, over the observational period.
 \begin{table}[t]
\normalsize
\caption{Systematic error in $R_0$ because of
neglect of a possible radial motion of Sgr~A*}
\label{tab_r0sys}
\vskip 0.01\textheight
\begin{center}
\renewcommand{\arraystretch}{1.2}
\begin{tabular}{lccccc}
\hline
Observational & $\langle V_r\rangle$ & $V_r(\text{Sgr A*}) $ &  $\delta_{\text{sys}}$ & \multicolumn{2}{c}{$\sigma_{\text{sys}}(R_0)$ (kpc)}\\
\cline{5-6}
Period       & (km/s)             & (km/s)             &                        & $R_0=7.5$ kpc & $R_0=8$ kpc  \\
\hline
2003 April--June & $-1500$ & 20 & 0.013 & 0.10 & 0.11 \\
                 &         & 60 & 0.040 & 0.30 & 0.32 \\
2004 July--August& $-1075$ & 20 & 0.019 & 0.14 & 0.15 \\
                 &         & 60 & 0.056 & 0.42 & 0.45 \\
\hline
 \end{tabular}
\end{center}
\end{table}           

Table~\ref{tab_r0sys} demonstrates that {\em neglect of even moderately
low   radial velocity of the orbital focus (Sgr~A*)  relative to the LSR
can lead to a substantial systematic error in $R_0$\/}: values of
$V_r^*=20$--60~km/s result in systematic $R_0$ errors of 1.3--5.6\%,
i.e., {\bf (0.1--0.45)$\mathbf{\times (R_0/8)}$~kpc}\/, for current
typical star's velocities. Notice that the value of
$\sigma_{\text{sys}}(R_0)$ can not be reduced statistically since {\em
all\/} $V_r$ values is biased coherently by any nonzero velocity  of
Sgr~A*. Only solving for $V_r(\text{Sgr A*}) $ can correct this
systematic error in $R_0$!

It should be mentioned that \cite{Tea06} state that they already solved 3D
velocity of Sgr~A$*$, however, not presenting in their short paper any
details---no  values of velocities and even no exact value of current point
estimate for $R_0$!

\subsection{Nonzero Proper Motion of Sgr~A*}

The reference frame for proper motions \citeauthor{Eea03}\ have
established by measuring the positions of nine astrometric reference
stars relative to typically 50--200 stars of the stellar cluster
surrounding Sgr A*; the uncertainty of the reference frame is
11.7~km/s \citep[see][]{Eea03}.  The effect of nonzero proper motion Sgr
A$^*$ relative to this frame, $\vec\mu^{\,*}\equiv\vec\mu(\text{Sgr A*})$, can be
approximately estimated if to imagine that the value of $R_0$ is
determined, also on the basis of $V_r$'s measurement at a moment
$t$, not from Eqs.~(\ref{V}) and (\ref{R0}) but from the ratio between
star's linear velocity on the sky, $V_\mu$, and star's proper motion,
$\mu$, measured for the same moment $t$:
\be\label{R0mu}
R_0=\frac{V_\mu}{\mu}.
\ee
The value of $V_\mu$ is a known function of $V_r$, orbital elements, and 
time:
\be\label{Vmu}
V_\mu^2=V^2-V_r^2=V_r^2(\Psi^{-2}-1),\qquad \Psi^2(t)\equiv \frac{V_r^2}{V^2},
\ee
where $\Psi^2(t)$ can be calculated from orbital elements [Eq.~(\ref{Vr/V})]. 
Any nonzero radial velocity $V_r^*$ and nonzero proper motion $\mu^*$ of
Sgr~A*  are equivalent to the introducing systematic errors
$\varepsilon_{V_\mu}$ and $\varepsilon_{\mu}$ in $V_\mu$ and $\mu$,
correspondingly. Because values of $V_r^*$ and $\mu^*$ are independent and
unknown, their combined impact on an $R_0$ estimate can be described by  the
formula of propagation of errors applied to Eq.~(\ref{R0mu}):
\bea
\varepsilon^2_{R_0}\equiv \sigma_{\text{sys}}^2(R_0)
	&=&
	\left(\frac{\varepsilon_{V_\mu}}{\mu}\right)^2+
	\left(\frac{V_\mu}{\mu^2}\varepsilon_\mu\right)^2\nonumber\\
        &=&(R_0/V_\mu)^2(\varepsilon^2_{V_\mu}+R_0^2\varepsilon^2_\mu).
\eea
From Eq.~(\ref{Vmu}) follows
\be
\varepsilon_{V_\mu}=\varepsilon_{V_r}\sqrt{\Psi^{-2}-1},
\ee
if an uncertainty on orbit elements is ignored, as it was actually done in
section~\ref{Vr_ne_0} Then considering that $\varepsilon_{V_r}=|{V_r^*}|$ we have
\be
\varepsilon^2_{R_0}=
        \frac{R_0^2}{V_r^2}\left({V_r^*}^2+R_0^2\varepsilon^2_\mu\frac{\Psi^2}{1-\Psi^2}\right).
\ee

Value of $\varepsilon_\mu$ depends from the relative orientation of vectors
$\vec\mu$ and $\vec\mu^{\,*}$. In the general case $0 \le \varepsilon_\mu \le
\mu^*$. Hence, e.g., for equal radial and tangential components of
Sgr~A* motion, i.e., for $V_\mu^*=|V_r^*|$, or $\mu^*=|V_r^*|/R_0$,
\be
\max\varepsilon_{R_0}=\varepsilon_{R_0}(V_r^*)k_1,\qquad k_1=\frac{1}{\sqrt{1-\Psi^2}},
\ee
\be
\varepsilon_{R_0}(V_r^*)\equiv R_0\left|\frac{V_r^*}{V_r}\right|.
\ee
Here $\varepsilon_{R_0}(V_r^*)$ is the systematic error in $R_0$ owing to
only the radial velocity of Sgr~A* [see Eq.~(\ref{delta})].

For ${V_\mu^*}^2=2{V_r^*}^2$, or $\mu^*=\sqrt{2}|V_r^*|/R_0$, i.e., for equal all
three Cartesian components of Sgr~A* motion,
\be
\max\varepsilon_{R_0}=\varepsilon_{R_0}(V_r^*)k_2,\qquad
k_2=\sqrt{\frac{1+\Psi^2}{1-\Psi^2}}.
\ee

With the S2 orbit elements derived in \cite{Eea05}, $k_1\approx 1.4974$,
$k_2\approx 1.8666$. 

Thus, for a given $V_r$ the effect of nonzero proper motion of Sgr A* on
$R_0$, being a function of the true anomaly, ranges from zero to values
comparable to the effect of nonzero radial velocity of Sgr~A*, in the latter
case increasing measurably the total systematic error in $R_0$.

\section{Conclusions}

Simple considerations show that {\em neglect of even low radial velocity
of Sgr~A* relative to the LSR leads to a substantial systematic error
in\/} $R_0$---up to 6\%, i.e., ${\sim}0.5$~kpc, for plausible values of
Sgr~A* velocity. It is too much to consider the distance to Sgr~A*, not
to mention the value of $R_0$, as being established reliable from the
present results on modelling the S2/Sgr~A* system.

A proper motion of Sgr~A* biases the distance value not so inevitably,
but in limiting cases can increase the systematic error in $R_0$ owing
to radial motion by factor up to ${\approx}1.5$--$1.9$ for similar
values of Sgr~A*'s tangential velocity.

\acknowledgments
I am grateful to Prof.~K.~V.~Kholshevnikov and to Prof.~S.~A.~Kutuzov
for valuable remarks and discussions. The work is partly supported by
the Russian Pre\-si\-dent Grant for State Support of Leading Scientific
Schools of Russia no.\ NSh-4929.2006.2.


\begin{thebibliography}{}

\bibitem[{Blitz(1994)Blitz}]%
	{Blitz94}	
	Blitz, L. 1994, in ASP Conf.\ Ser., Vol.~66, Physics of
                 the Gaseous and Stellar Disks of the Galaxy, ed.\
                 I. R. King (San Francisco: ASP), 1


\bibitem[{Eisenhauer et~al.(2005)Eisenhauer, Genzel, Alexander et~al.}]%
	{Eea05}	
	Eisenhauer, F., Genzel, R., Alexander, T., Abuter, R.,
	Paumard, T., Ott, T., Gilbert, A., Gillessen, S., Horrobin, M.,
        Trippe, S., Bonnet, H., Dumas, C., Hubin, N., Kaufer, A.,
        Kissler-Patig, M., Monnet, G., Str\"obele, S., Szeifert, T.,
        Eckart, A., Sch\"odel, R., \& Zucker,~S.
	2005, \apj, 628, 246

\bibitem[{Eisenhauer et~al.(2003)Eisenhauer, Sch\"odel, Genzel et~al.}]%
	{Eea03}
	Eisenhauer, F., Sch\"odel, R., Genzel, R., Ott, T.,
	Tecza, M., Abuter, R., Eckart, A., \& Alexander T.
        2003, \apj, 597, L121 

\bibitem[{Ghez et~al.(2003)Ghez, Duch\^ene, Vatthews et~al.}]%
	{Ghez_ea03}
	Ghez, A. M., Duch\^ene, G., Matthews, K., Hornstein, S. D., Tanner, A., 
	Larkin, J., Morris, M., Becklin, E. E., Salim, S., Kremenek, T., 
	Thompson, D., Soifer, B. T., Neugebauer, G., \& McLean, I.
	2003, \apj, 586, L127

\bibitem[{Mouawad et~al.(2003)Mouawad, Eckart, Pfalzner et~al.}]%
	{Mouawad_ea05}
	Mouawad, N., Eckart, A., Pfalzner, S., Sch\"odel, R., Moultaka, J., \& Spurzem, R.
        2005, Ast.\ Nachr., 326, 83 

\bibitem[{Nikiforov(2004)Nikiforov}]%
	{N04}
	Nikiforov, I. I., 2004, in Order and Chaos in Stellar and Pla\-ne\-tary
        Systems, ed.\ G. G. Byrd, K. V. Kholshevnikov,  A. A. Myll\"ari, {\em
	et al.}, ASP Conf.\ Ser.\ (San Francisco: ASP),
	316, 199

\bibitem[{Nishiyama et~al.(2006)Nishiyama, Nagata, Sato et~al.}]%
	{Nishiyama_ea06}
	Nishiyama, S., Nagata, T., Sato, S., Kato, D., Nagayama, T., Kusakabe, N., 
	Matsunaga, N., Naoi, T., Sugitani, K., \& Tamura, M.
	2006, \apj, 647, 1093

\bibitem[Reid(1993)]{R93} Reid, M. J. 1993, \araa, 31, 345

\bibitem[{Sch\"odel et~al.(2002)Sch\"odel, Ott, Genzel et~al.}]%
	{Schoedel_ea02}
	Sch\"odel, R., Ott, T., Genzel, R., Hofmann, R., Lehnert, M., Eckart, A., 
	Mouawad, N., Alexander, T., Reid, M. J., Lenzen, R., Hartung, M., 
	Lacombe, F., Rouan, D., Gendron, E., Rousset, G., Lagrange, A.-M., 
	Brandner, W., Ageorges, N., Lidman, C., Moorwood, A. F. M,. Spyromilio, J., 
	Hubin, N., \& Menten, K. M. 2002, Nature, 419, 694

\bibitem[{Subbotin(1968)Subbotin}]%
	{Subbotin68}	
	Subbotin, M. F. 1968, Introduction in Theoretical Astronomy 
	(Moscow: Nauka),  800~pp. (in Russian)

\bibitem[{Trippe et~al.(2006)Trippe, Gillessen, Ott et~al.}]%
	{Tea06}
	Trippe, S., Gillessen, S., Ott, T., Eisenhauer, F., Paumard, T., 
	Martins, F., Genzel, R., Sch\"odel, R., Eckart, A., \& Alexander, T.
	2006, Journal of Physics: Conf.\ Ser., 54, 288

\bibitem[{Weinberg et~al.(2005)Weinberg, Milosavljevi\'c, \& Ghez}]%
	{Weinberg_ea05}
	Weinberg, N.\ N., Milosavljevi\'c M., \& Ghez, A.\ M.
	2005, in ASP Conf.\ Ser., Vol.~338, Astrometry in the Age of the Next 
	Generation of Large Telescopes, ed.\ P.\ K.\ Seidelmann \& A.\ K.\ B.\ Monet 
	(San Francisco: ASP), 252

\end{thebibliography}
\end{document}